\documentstyle[twoside,epsfig,11pt]{elsart}
\journal{Astroparticle Physics}
\begin{document}
\begin{frontmatter}
\title{An Upper Limit on the Infrared Background Density from HEGRA data on Mkn\,501\thanksref{BMBF}}
\thanks[BMBF]{This work is supported by the BMBF, FRG, under contract
number 05\,2\,WT\,164.}

\author{B.\,Funk},
\author{N.\,Magnussen\thanksref{email}}, 
\author{H.\,Meyer},
\author{W.\,Rhode},
\author{S.\,Westerhoff\thanksref{now}},
\author{B.\,Wiebel-Sooth}
\address{Universit\"at Wuppertal, Fachbereich Physik,
Gau\ss str.20, 42097 Wuppertal, Germany}

\thanks[email]{corresponding author: magnus@wpos7.physik.uni-wuppertal.de}
\thanks[now]{Now at: University of California, Santa Cruz, CA 96064, USA}

\begin{abstract}
The energy spectrum of Mkn\,501 in the TeV energy regime, as measured by the
HEGRA (High Energy Gamma Ray Astronomy) \v{C}erenkov telescopes during
its low state in 1995/96 and during a fraction of the 1997 outburst in the TeV 
energy regime, is shown to place stringent upper limits on the still unknown
infrared photon density in the energy region between 3$\cdot 10^{-3}$ and $3\cdot 10^{-1}$
eV. Assuming two different shapes for the unknown infrared photon spectrum
in this energy range we calculate upper limits on the infrared photon density
on the basis of the power-law fit obtained for
the observed spectrum up to the maximum energy.
\end{abstract}
\end{frontmatter}

\newpage
\section{Introduction}

The \hfill cosmologically \hfill important \hfill extragalactical
\hfill diffuse \hfill
infrared \hfill background 
\newline (DIRB) in the astronomical window
from the optical to the infrared (IR)
has not yet been directly determined experimentally due to
large sytematic errors driven
by local effects. As realized
already a long time ago, 
the detection of extragalactic TeV-$\gamma$ sources
would enable one to indirectly measure this photon density due
to unavoidable pair production losses of TeV photons in
infrared photon fields 
(Gould \& Schr\'eder 1966, Stecker et al. 1992). 
Interpreting the Whipple data of Mkn\,421 (redshift $z$ = 0.031), Stecker et al. (1994)
claimed to have seen an exponential cut-off
resulting from cosmic absorption 
of an otherwise smooth power law $\gamma$-ray spectrum.
The authors suggested that the IR-density is given by 
$n(\epsilon)\approx (0.8^{+0.6}_{-0.4})\cdot 10^{-3}\epsilon^{-2.6}(h/0.75)$
cm$^{-3}$eV$^{-1}$
where $h$ refers to the normalization factor of the Hubble constant
($H_\circ = h 100$km s$^{-1}$ Mpc$^{-1}$). A
similar analysis has been carried out by Dwek \& Slavin (Dwek \& Slavin 1994).
The inferred large values of the diffuse near-infrared background density
extrapolated to the infrared would imply that it is virtually impossible
to discover any extragalactic source above a few TeV.
However, Biller et al. (Biller et al. 1995) correctly pointed out that 
unless the source spectrum is known, one measured TeV source can only yield
an upper limit, because the cut-off may be due to internal absorption at the
source. They obtain a conservative upper limit of $\epsilon^2 n(\epsilon)=
0.04$\,eV\,cm$^{-3}$ at $\epsilon=0.1$\,eV. \\
In the meantime, besides Mkn\,421 (Punch et al. 1992, Petry et al. 1996)
Mkn\,501 (redshift $z$ = 0.034) is now the
second extragalactic TeV source to be
discovered and extensively monitored  
with the Whipple and HEGRA \v{C}erenkov
telescopes (Kerrick et al. 1995a, Bradbury et al. 1997).
Both objects belong to the blazar subclass of 
galaxies showing powerful non-stellar activity characterized by
rapidly variable, polarized continuum emission. 
As the outburst of Mkn\,501 in early March 1997 
(Aharonian et al. 1997) and
the two outbursts of Mkn\,421 in Spring of 1995 and 1996 
(Kerrick et al. 1995b, Gaidos et al. 1996, Buckley et al. 1996) have shown, both objects
also exhibit rapid variability with large amplitudes in the TeV energy range.
The HEGRA collaboration showed that 
between March 16 and March 20 (about 27 hours observation
time) the Mkn\,501
energy spectrum in the TeV energy range
extended beyond 10 TeV without a visible break in the power law
spectrum with a differential power 
law index of 2.49 $\pm$ 0.11 (stat.) $\pm$ 0.25 (syst.) (Aharonian et al. 1997).
During the quiescent state of Mkn\,501 in 1996, this source was 
observed with the HEGRA CT1 telescope for 220 hours. 
The measured energy spectrum for this
period can be described by a differential power
law index of 2.5 $\pm$ 0.4 (total error) (Petry 1997). 
\\
In the following we first discuss the current situation
regarding models and observations of the DIRB and then
derive an upper limit on the DIRB from the measured energy spectrum
of Mkn\,501 in 1997 for IR photon energies between 3$\cdot 10^{-3}$ eV and
3$\cdot 10^{-1}$ eV. 

\section{Models and Observations of the DIRB}
\begin{figure}[ht]
\begin{center}
\epsfig{file=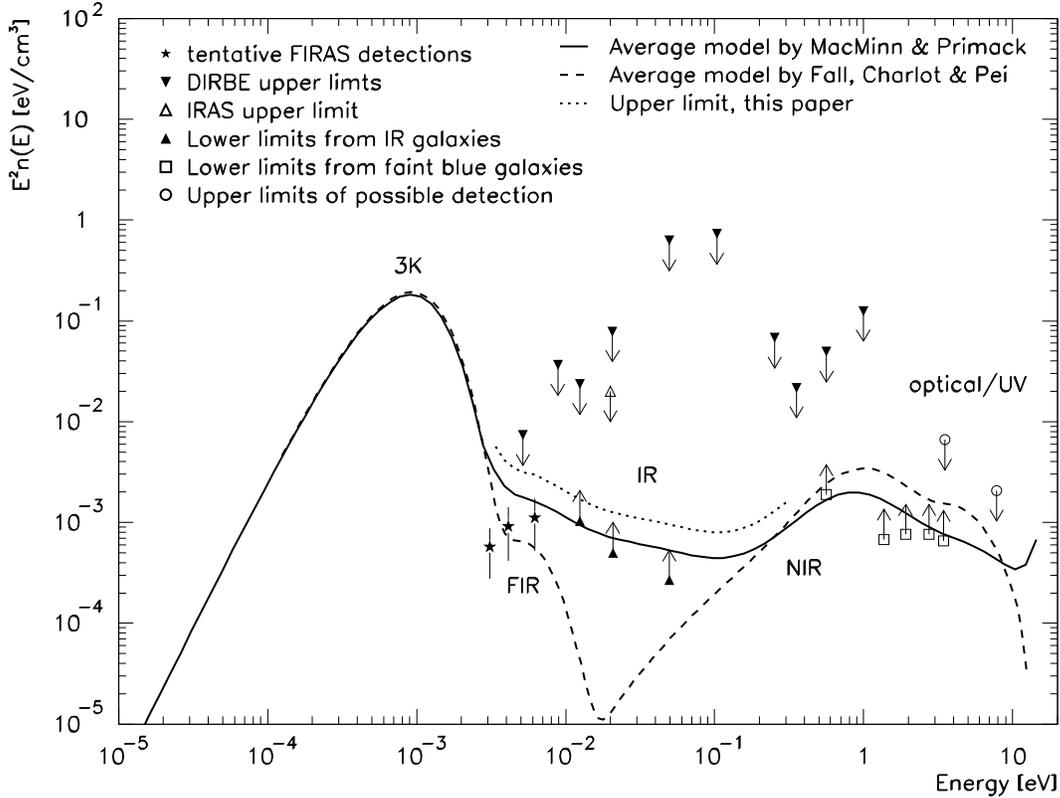,height=10.5cm}
\caption[Energy density of the extragalactic diffuse background radiation]{
\label{dirb}Energy density of the extragalactic diffuse background radiation;
{\it stars:} tentative FIRAS detection of the FIR background without CMBR
(Puget et al. 1996), {\it full downward triangles:} 
upper limits from the DIRBE experiment ([Hauser, p. 135], all 
references in square brackets are taken from Calzetti et al. 1995),
{\it squares:} lower limit from faint blue galaxies [Tyson, p. 103],
{\it circles:} upper limit of possible detection [Paresce, p. 307],
{\it open triangle:} IRAS upper limit (Boulanger \& Perault 1988),
{\it full upward triangles:} evolution model
dependent lower limits from number counts of
infrared-bright galaxies from Hacking \& Soifer 1991,
{\it solid line:} average model from MacMinn \& Primack including
the CMBR (MacMinn \& Primack 1996),
{\it dashed line:} average model from Fall, Charlot \& Pei added to the CMBR 
(Fall et al. 1996), {\it dotted line:} upper limit derived in this paper.}
\end{center}
\end{figure}

The current experimental and model situation regarding the DIRB is summarized
in fig.~\ref{dirb}. Since the published energy spectra for both of the
extragalactic sources show no spectral break we do not
incorporate the claimed detections of the DIRB density based
on preliminary Mkn\,421 spectra (Stecker et al. 1994).
We include, however, the recent tentative detection of the
far-infrared background radiation (Puget et al. 1996) by COBE which
is an important 
step towards a direct measurement of the flux in the IR-regime
and which is at the moment
rather weakly constrained by the results given by Hauser
in Calzetti et al. (1995). \\
Due to the cosmological implications of the DIRB photon density
a great number of
diffuse IR models ranging from simple power laws to multicomponent spectra
have been developed by many authors.
A large number of parameters 
such as the star/galaxy formation rate, the
number distribution of star masses, the dust content and all cosmological
parameters enter into these models. In fig.~\ref{dirb} 
we show the predictions of two complex models: The model 
which we
later employ
for the numerical calculation
is by MacMinn \& Primack (MacMinn \& Primack 1996) who have
calculated the DIRB for various realizations of cosmological parameters and
dark matter models. The results clearly
depend on the set of parameters, and the flux predictions vary at most by a
factor of 3. Accounting for the uncertainties of the model an
average DIRB spectrum is assumed in the following.
The model by Fall, Charlot \& Pei (Fall et al. 1996)
is based on the assumption that the star formation rate is directly
related to the consumption of neutral gas. 
The density of neutral gas can
be determined from analyses of Ly$\alpha$ absorption lines as seen in distant
quasar spectra thus leading to a cosmological rate of star formation out
to distances of $z\simeq 5$. Once the rate is known, the emissivity of the
universe in the IR can be calculated as a function of $z$. The dashed curve
in fig.~\ref{dirb} shows
the case where the neutral gas is fully consumed during the formation 
process referred to as the closed box scenario. 

\section{Gamma-ray absorption}
\begin{figure}
\centerline{\epsfig{figure=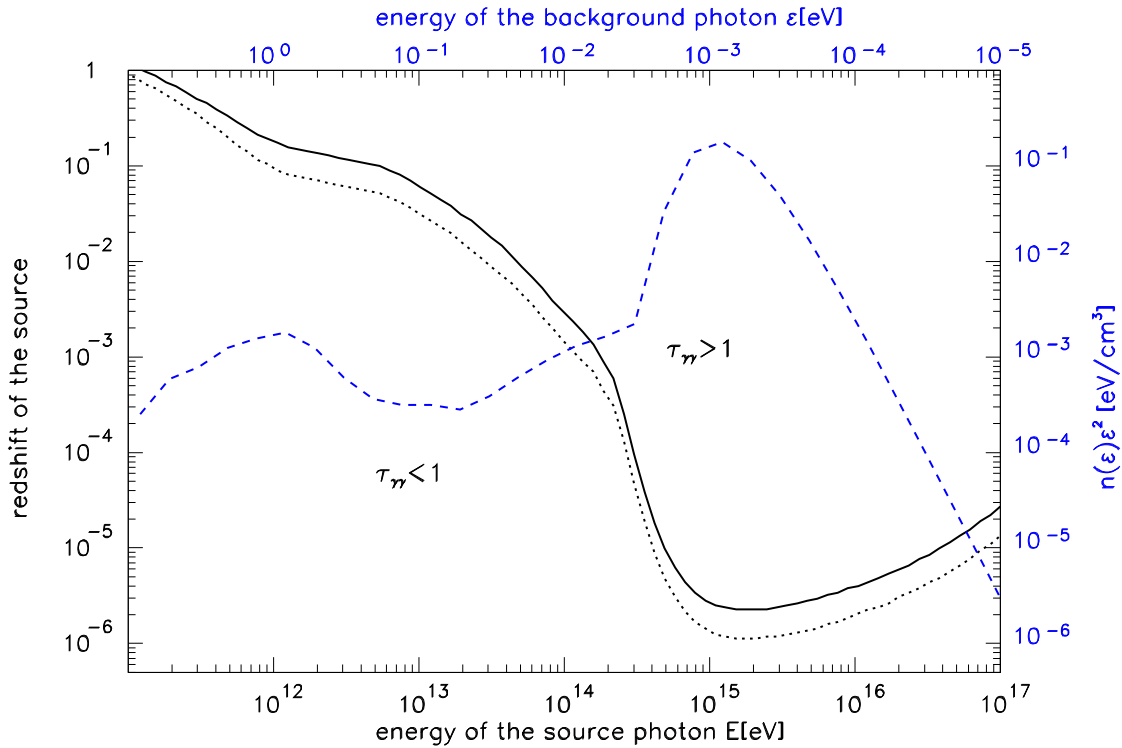,height=10cm,width=15cm}}
\caption{\label{ghorizon} $\gamma$-ray horizon for the averaged 
diffuse background radiation model from MacMinn \& Primack (fig.~\ref{dirb})
{\it Solid line: $h=1.0$. Dotted line:  $h=0.5$}. To guide the eye the
DIRB model has been added to the figure with an inverse energy scale.}
\end{figure}

For $\gamma$-rays
of energy $E$ propagating from a distant source at redshift
$z_\circ$ towards a terrestrial observer the threshold energy for
pair creation in
interactions with low energy photons of present-day energy $\epsilon$
from an isotropic diffuse background radiation field
is given by
\begin{equation}
\label{eps}
\epsilon_{\rm th}={2(m_{\rm e}c^2)^2\over E(1-\mu)(1+z_{\circ})^2}
\end{equation}
where $\mu=\cos\theta$ denotes the cosine of the scattering angle. 
A 
soft photon density strongly varying with energy is thus reflected
in the
optical depth $\tau_{\gamma\gamma}$ determining the
number density of target photons at the resonant energy $\propto E^{-1}$.
The pair creation cross section is given by
\begin{equation}
\label{sigma}
\sigma_{\gamma\gamma}={3\sigma_{\rm T}\over 16}
(1-\beta^2)\left[2\beta(\beta^2-2)+(3-\beta^4)\ln\left(1+\beta\over
1-\beta\right)\right]
\end{equation}
with
\begin{equation}
\beta=\sqrt{1-{1\over \gamma^2}} \quad\mbox{and}\quad \gamma^2=
{\epsilon\over\epsilon_{\rm th}}.
\end{equation}
Here $\sigma_{\rm T}=6.65\cdot 10^{-25}$\,cm$^2$ denotes the Thomson cross
section. 
For the computation of the optical depth we use the geodesic radial
displacement function $dl/dz={c\over H_\circ}[(1+z)E(z)]^{-1}$.
With the proper physical distance, $l(t)$, between a pair of well-separated points
as a function of time
given by $l(t) = l_0 a(t)$, the cosmological 
expansion rate is $H(t) = {\dot{a}\over a} = H_\circ E(z)$
with the
dimensionsless function $E(z)$.
Under the assumption that the
mean mass density is dominated by non-relativistic matter
the cosmological equation for the expansion rate is given by 
(see e.\,g.\,Peebles 1993, eq. (13.3))
\begin{equation}
{\dot{a}\over a} = H_\circ E(z) = 
H_\circ \lbrack \Omega (1+z)^3 + \Omega_R (1+z)^2 + \Omega_\Lambda \rbrack^{1/2}
\vspace*{-0.3cm}
\end{equation}
\noindent
where $\Omega, \Omega_R,$ and $\Omega_\Lambda$ 
are the three contributions to the Hubble constant
due to the present mean mass density, 
the radius of space curvature, and the cosmological
constant $\Lambda$, respectively, with $\Omega +\Omega_R + \Omega_\Lambda = 1$.
For a cosmological model with
$\Omega=1$, $\Lambda=0$, and negligible space curvature
the function $E(z)$ simplifies to $(1+z)^{3/2}$
and the optical depth can be written as
\begin{eqnarray}
\label{taueq}
\tau_{\gamma\gamma}(E,z_\circ)=
\int_0^{z_\circ}dz{dl\over dz}\int_{-1}^{+1}d\mu{1-\mu\over 2}
\int_{\epsilon_{\rm th}}^\infty
d\epsilon \, n_{\rm b}(\epsilon)(1+z)^3\sigma_{\gamma\gamma}(E,\epsilon,\mu,z)\\
={c\over H_\circ}\int_0^{z_\circ}dz
(1+z)^{1/2}\int_{0}^{2}d\mu {\mu\over 2}\int_{\epsilon_{\rm th}}^\infty
d\epsilon \, n_{\rm b}(\epsilon)\sigma_{\gamma\gamma}(E,\epsilon,\mu,z)
\end{eqnarray}
for a non-evolving present-day background density $n_{\rm b}$, i.e.
$n_{\rm b}'(z,\epsilon')d\epsilon'=(1+z)^3n_{\rm b}(\epsilon)d\epsilon$
where the prime indicates comoving frame quantities. Numerical results
for the optical depth using specific models for the background radiation
will be obtained below.\\
Under the assumption of a particular model for the IR-density
one can numerically integrate equation (\ref{taueq}) and obtain the
optical depth $\tau_{\gamma\gamma}(E,z)$. 
For a source spectrum $\Phi(E)$
the observed spectrum is then simply 
given by $\Phi(E)\times \exp(-\tau_{\gamma\gamma}(E,z))$.
While
for an optical depth smaller than unity the universe appears transparent,
it becomes opaque for larger values of $\tau_{\gamma\gamma}(E,z)$.
This defines the $\gamma$-ray horizon ($\tau_{\gamma\gamma}(E,z)=1$) which
is the size of the visible $\gamma$-ray universe at a given energy.
The $\gamma$-ray horizon is shown for two values of the Hubble constant
in fig.~\ref{ghorizon}.
It turns out that at energies between 10$^{10}$ and 10$^{15}$\,eV
the size of the visible universe is a decreasing function of energy.
Note that the adopted model does not take into account the extragalactic
diffuse radio background which only becomes
important above the characteristic energies of air showers
accessible by HEGRA-type experiments. 

\section{Results}
\begin{figure}
\centerline{\epsfig{figure=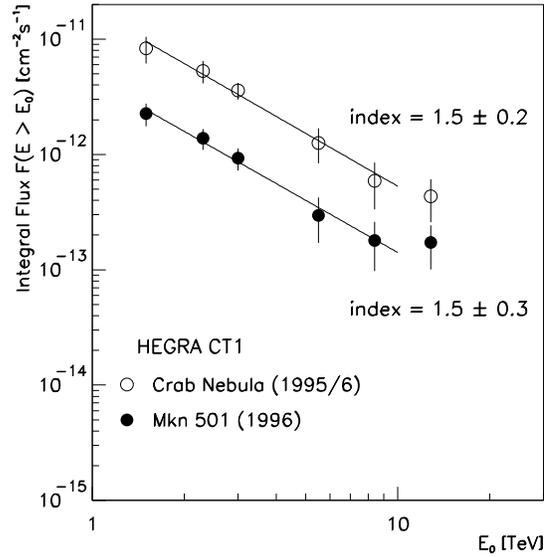,height=9cm,width=9cm}}
\caption{\label{spectrum1}
Average integral spectrum of $\gamma$-rays from Mkn 501 as
measured in 1996 (220 hours)
and from the Crab nebula as measured in the period 1995/96 (37 hours)
with the HEGRA CT1 telescope.
The lines
represent power-law fits. The error bars correspond to the statistical errors
(from Petry 1997).
}
\end{figure}

\begin{figure}
\centerline{\epsfig{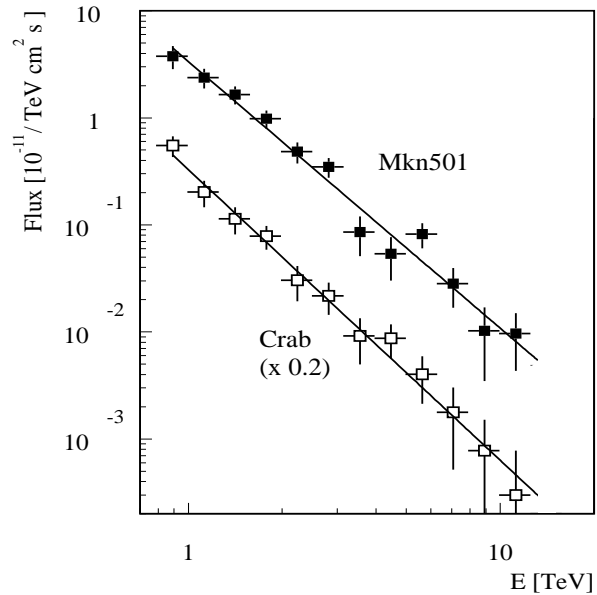}}
\caption{\label{spectrum2}
Average differential spectrum of $\gamma$-rays from Mkn 501 between
March 15 and March 20, 1997 (27 hours) and from the Crab nebula (10
hours) as measured with the
HEGRA CT system. The Crab data points are scaled by a factor of 0.2. The lines
represent power-law fits. Only statistical errors are shown. The energy scale
has a 20\% systematic error (from Aharonian et al. 1997).
}
\end{figure}

\begin{figure}
\centerline{\epsfig{figure=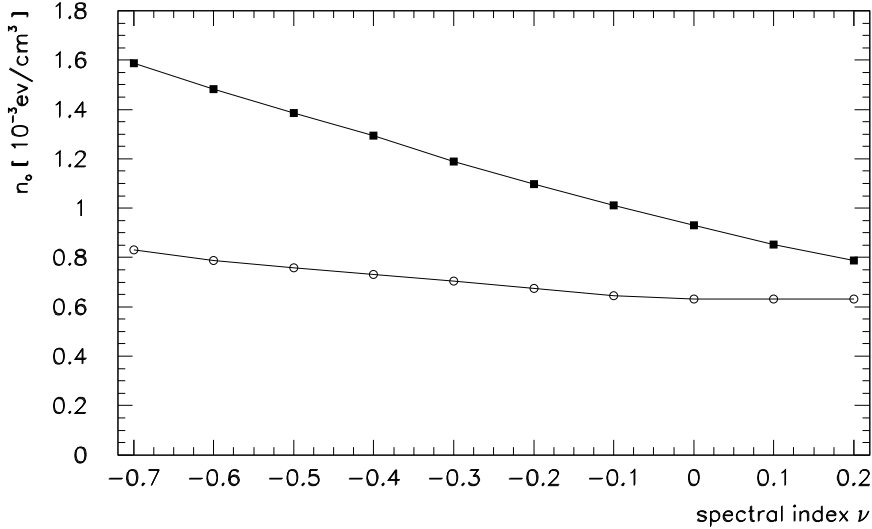,height=7cm,width=12cm}}
\caption{\label{irexcl} Results from this analysis: Values for the spectral index
$\nu$ and $n_\circ$ above the upper curve are forbidden given the
unattenuated observation of Mkn\,501
at 10\,TeV ({\it squares}). For illustration purposes we add
the exclusion curve that would result {\it if} no spectral break
would be observed up to 15\,TeV ({\it circles}).}
\end{figure}

\begin{figure}[ht]
\begin{center}
\epsfig{file=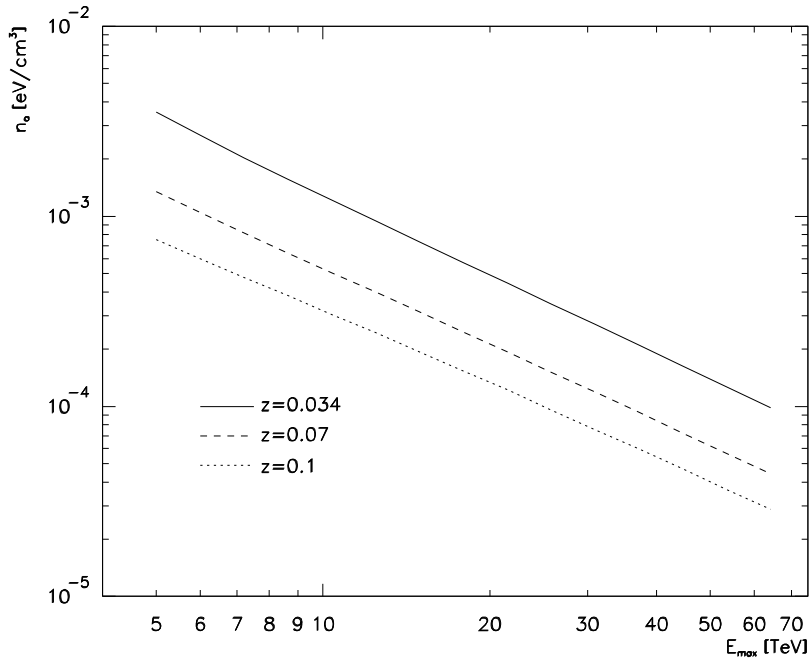,height=8.5cm}
\caption[Maximum observable energy.]{
\label{irul}The normalization $n_0$ of the infrared photon density
in the energy range between $3 \cdot 10^{-3}$ to  $3 \cdot 10^{-1}$\,eV
assuming a spectrum
$\epsilon^2 n(\epsilon) = n_0(\epsilon/3 \cdot 10^{-2}\,{\rm
eV})^{-0.4}$ which would result from our calculations as a function of the
maximum observable energy (with optical depth $\tau$ =
1) for three values of redshift $z$.}
\end{center}
\end{figure}

Petry (Petry 1997) and 
Aharonian et al. (Aharonian et al. 1997) have
shown that the TeV energy
spectrum of Mkn\,501 can be described by an unattenuated
power law with differential indices
of $-2.5\pm 0.4$ (total error) in 1996 and $-2.49\pm 0.11$ (stat.)
in May 1997, respectively.
For the analysis of the 1996 data taken with the HEGRA CT1 telescope
during about 220 hours of observation time in fig.~\ref{spectrum1} we show
the integral flux spectrum
with the fitted power law index of  $-1.5\pm 0.3$ (total error)
indicating the reduced error in
the fit of the integral spectrum due to the correlated bins.
This flux is compared to the Crab nebula integral spectrum as measured
in the 1995/96 observation period with the CT1 telescope (37 hours).
From the analysis of the 1997 Mkn\,501 (27 hours)
and Crab nebula data (10 hours) as measured 
with the HEGRA CT system
we show the differential
flux spectra in fig.~\ref{spectrum2}.
From the observed unattenuated energy spectrum
extending up to 10\,TeV we conclude that 
the optical depth for this source is less than unity, i.\,e.\,,
$\tau_{\gamma\gamma}(E={\rm 10\,TeV},z=0.034)<1$. To derive the upper limit 
on the DIRB photon density we
take $\tau_{\gamma\gamma}=1.0$
and $h=1.0$, and assume a specific shape of the IR-density.\\\
For the energy range $3\cdot 10^{-3}\,$eV $< \epsilon <3\cdot 
10^{-1}$\,eV we make an ansatz
for the infrared spectrum based on the
results of two different models (outside this energy range only the MacMinn \& 
Primack model shown in fig.~\ref{dirb} is used): 
(i) $\epsilon^2 n(\epsilon) = n_\circ (\epsilon/E_p)^\nu$ with
the pivot energy point $E_p = 3\cdot 10^{-2}\,{\rm eV}$
and (ii) the shape of the spectrum in this energy range is assumed to
be identical with the models by
MacMinn \& Primack but its absolute level is
varied in the calculation. The results for
the first ansatz are presented in fig.~\ref{irexcl} where values of $\nu$ and
$n_\circ$ below the curve are still allowed by the detection of Mkn\,501 at 
10\,TeV. To illustrate the impact of a still not excluded
observation of the unabsorbed spectrum extending up to 15\,TeV,
a second curve was added in the figure. 
Using the second model ansatz the observation of Mkn\,501 at 10\,TeV
restricts the normalization $N$  of the spectral segment between 
$3\cdot 10^{-3}$\,eV and $3\cdot10^{-1}$\,eV to $N<1.8$ (times the flux
assumed in this work). At the pivot energy point of $3 \cdot
10^{-2}$\,eV we thus determine upper limits of
$\epsilon^2 n(\epsilon) = 1.1 \cdot 10^{-3}$\,eV/cm$^3$
for the MacMinn \& Primack model and $1.0  \cdot 10^{-3}$\,eV/cm$^3$
for the powerlaw ansatz and assuming the slope of the spectrum
to be flat ($\nu = 0$) around the pivot point.
A comparison with fig.~\ref{dirb}, where we have added the
upper limit curve derived using the MacMinn \& Primack ansatz, indicates 
that these limits are
about two orders of magnitude below those accessible 
by direct measurements at this energy and that they are compatible 
with the tentative FIRAS measurement of the IR density (Puget et al. 1996).  \\
As pointed out in the introduction, the measurement of single spectra,
even if a break-off feature is observed, does not permit
a determination of the actual
level of the DIRB density. Up to the end of the 1997 observation period (October 1997)
the blazar Mkn\,501 continued to be in a high state with respect to
TeV emission. The increasing statistics will enable
\v{C}erenkov telescope experiments to extend observations
to higher energies and search for cut-off features.
The relationship between the infrared photon density
in the energy range from  $3 \cdot 10^{-3}$ eV to  $3 \cdot 10^{-1}$\,eV,
assuming a powerlaw spectrum 
$\epsilon^2 n(\epsilon) = n_0(\epsilon/3 \cdot 10^{-2}\,{\rm
eV})^{-0.4}$, and the maximum observable energy, i.\,e., 
$E_{max}$ with optical depth $\tau$ = 1, 
for different values of redshift $z$ is shown in fig.~\ref{irul}.
If e.\,g.\,no
cut-off is observed up to around 70\,TeV we would thus derive,  
using the above described procedure, an upper limit on the DIRB density
of about 0.1 times the average MacMinn \& Primack model prediction. 

\section{Conclusions}
Based on the observation
of the {\sl unabsorbed} Mkn\,501 $\gamma$-spectrum extending beyond 10\,TeV,
we have derived a stringent upper limit on the 
extragalactic diffuse infrared photon energy density in the energy range
from $3 \cdot 10^{-3}$ to $3 \cdot 10^{-1}$\,eV of 1.8 times 
the prediction of an average model by MacMinn \& Primack.
This translates into an upper 
limit of the energy density at the used pivot energy point of $3 \cdot
10^{-2}$\,eV  of
$\epsilon^2 n(\epsilon) = 1.1 \cdot 10^{-3}$\,eV/cm$^3$.
For the second ansatz, i.\,e.\,,
a power-law ansatz for the DIRB around this pivot energy point,
we determined upper limits on the normalization as a function of the
spectral index, e.\,g.\,for a flat
spectrum $\epsilon^2 n(\epsilon) = n_0$, the resulting upper limit is
$n_0 = 1.0 \cdot 10^{-3}$\,eV/cm$^3$.
These upper limits are about 2 orders of magnitude {\sl below} upper limits
derived from 
current direct measurements at this energy and are not in
contradiction to a value of the DIRB density
derived from preliminary evidence of TeV emission
reported by Meyer \& Westerhoff ( Meyer \& Westerhoff 1996). \\
The results presented in this paper are well compatible with other
analyses of the infrared photon density based 
on the HEGRA CT data of Mkn\,501 (Mannheim 1997, Stanev \& Franceschini 1997).
They are also in good agreement with an empirical calculation
of the DIRB based on galaxy luminosity functions in the IR (Malkan \& Stecker
1997) which in turn is in good agreement with recent DIRB models 
derived from star formation data (Guiderdoni et al. 1997).

\begin{ack}
We thank K.\,Mannheim, G\"ottingen, for useful discussions
and R.\,S.\,Miller, Los Alamos, for helpful comments on the manuscript.
\end{ack}


\begin{thebibliography}{}

\bibitem{}
Aharonian F.A. et al.,1997, submitted to A\&A

\bibitem{}
Biller S.D., Akerlof C.W., Buckley J.H, et al., 1995, ApJ 445, 227

\bibitem{}
Boulanger F. and Perault M., 1988, ApJ 330, 964.

\bibitem{}
Bradbury S.M., Deckers T., Petry D., et al., 1997, A\&A 320,  L5

\bibitem{}
Buckley J.H., et al., 1996, ApJ 472, L9

\bibitem{}
Calzetti D., Livio M., Madau P. (eds), 1995, {\sl Extragalactic
background radiation}, Space Telescope Science Institute Symposium,
vol.7, Cambridge Univ. Press

\bibitem{}
Dwek E. and Slavin J., 1994, ApJ 436, 696

\bibitem{}
Fall S.M., Charlot S., and Pei Y.C., 1996, ApJ 464, L43

\bibitem{}
Gaidos J.A., et al., 1996, Nature 383, 319

\bibitem{}
Gould R.\,J. and Schr\'eder G., 1966, Phys.\,Rev.\,Lett. 16, 252

\bibitem{}
Guiderdoni B. et al., 1997, Nature 390, 257

\bibitem{}
Hacking P.B. and Soifer B.T., 1991, ApJ 367, L49

\bibitem{}
Kerrick A.\,D. et al. , 1995a, ApJ 452, L59

\bibitem{}
Kerrick A.\,D., et al., 1995b, ApJ 438, L59

\bibitem{}
MacMinn D. and Primack J.R., 1996, Spac. Sci. Rev. 75, no.1-2, 413

\bibitem{}
Malkan M.\,A. and Stecker F.\,W., 1997, ApJ {\sl in press},
astro-ph/9710072

\bibitem{}
Mannheim K., 1997, Science, {\sl to appear}

\bibitem{}
Meyer H. and Westerhoff S., 1996, Proceedings of the Heidelberg
Workshop on {\sl GAMMA-RAY EMITTING AGN}, Heidelberg, Germany

\bibitem{}
Peebles P.J.E., 1993, {\sl Principles of Physical Cosmology}, Princeton
Univ. Press

\bibitem{}
Petry D., Bradbury S.M., Konopelko A., et al., 1996, A\&A 311, L13

\bibitem{}
Petry D., 1997, PhD thesis, MPI f\"ur Physik, M\"unchen (MPI-PhE/97-27)

\bibitem{}
Puget J.L., Abergel A., Bernard J.P., et al., 1996, A\&A 308, L5

\bibitem{}
Punch C.W., et al., 1992, Nature 358, 477

\bibitem{}
Stanev T. and Franceschini A., 1997, astro-ph/9708162

\bibitem{}
Stecker F.\,W., de Jager O.\,C., Salamon M.\,H., 1992, ApJ 390, L49

\bibitem{}
Stecker F.\,W., de Jager O.\,C., Salamon M.\,H., 1994,
Nature 369, 294

\end{thebibliography}
\end{document}